\documentclass[11pt,twoside]{article} 
\usepackage{cspm-asp2006}
\usepackage{epsfig,graphicx,natbib,url}  
\usepackage{lscape} 
\begin{document}

\markboth{Asensio Ramos and Trujillo Bueno}{Code for Chromospheric Plasma} 
\title{A User-Friendly Code to Diagnose Chromospheric Plasmas} 
\author{A. Asensio Ramos$^{1}$ and J. Trujillo Bueno$^{1,2}$} 
\affil{$^1$Instituto de Astrof\'{\i}sica de Canarias, La Laguna, Spain \\
       $^2$Consejo Superior de Investigaciones Cient\'{\i}ficas, Spain} 
  
\begin{abstract} 
The physical interpretation of spectropolarimetric observations of
lines of neutral helium, such as those of the 10830\,\AA\ multiplet,
represents an excellent opportunity for investigating the magnetism of
plasma structures in the solar chromosphere. Here we present a
powerful forward modeling and inversion code that permits either to
calculate the emergent intensity and polarization for any given
magnetic field vector or to infer the dynamical and magnetic
properties from the observed Stokes profiles. This diagnostic tool is
based on the quantum theory of spectral line polarization, which
self-consistently accounts for the Hanle and Zeeman effects in the
most general case of the incomplete Paschen-Back effect regime. We
also take into account radiative transfer effects. An efficient
numerical scheme based on global optimization methods has been
applied. Our Stokes inversion code permits a fast and reliable
determination of the global minimum.
\end{abstract}

\section{Introduction}
The quality of spectropolarimetric observations of solar chromospheric
plasmas is steadily increasing. However, in order to obtain reliable
empirical information on the strength and geometry of the magnetic
field vector we also need to develop and apply suitable diagnostic
tools within the framework of the quantum theory of spectral line
polarization \citep[e.g., the recent monograph
by][]{aar-landi_landolfi04}. In particular, the development of a
forward modeling and inversion code of Stokes profiles is of great
importance because it would facilitate the investigation of a variety
of problems of great astrophysical interest. For such a diagnostic
tool to be reliable, it has to include all the important physical
mechanisms that induce spectral line polarization in stellar
atmospheres \citep[e.g., the recent review
by][]{aar-trujillo_lybia06}. Not only the Zeeman effect, but also the
atomic level polarization induced by anisotropic radiation pumping and
its modification by the Hanle effect have to be taken into
account. For a remarkable example of a pioneering work in this field
see Landi Degl'Innocenti (1982).

One of the main motivations of our investigation has been to develop a
robust but user-friendly computer program for facilitating the
analysis of the polarization signals induced by the above-mentioned
physical mechanisms in the spectral lines of the He\,\uppercase{i}
10830\,\AA\ multiplet, which are sensitive to both the Zeeman effect
and to the presence of atomic level polarization
(\cite{aar-trujillo_nature02};
\cite{aar-trujillo_helium07}). Spectro-polarimetric observations of
this multiplet have been carried out in a variety of plasma structures
of the solar chromosphere and corona, such as sunpots
\citep{aar-harvey_hall71,aar-ruedi95,aar-centeno06}, coronal filaments
\citep{aar-lin98,aar-trujillo_nature02}, prominences
\citep{aar-trujillo_nature02,aar-merenda06}, emerging magnetic flux
regions \citep{aar-solanki_nature03,aar-Lagg04}, chromospheric
spicules \citep{aar-trujillo_merenda05}, active region filaments
(Mart\'\i nez Pillet et al.\ 2007; in preparation) and flaring regions
(Sasso et al.\ on p.~467\,ff of these proceedings).

\section{The Forward Modeling Option}
The adopted atomic model includes the following five terms of the
triplet system of He\,I: 2s$^3$S, 2p$^3$P, 3s$^3$S, 3s$^3$P and
3d$^3$D. It has been concluded that this 11 $J$-levels model with 6
radiative transitions between the terms is enough for a satisfactory
modeling of the polarization properties of the spectral lines of the
He\,\uppercase{i} D$_3$ multiplet at 5876\,\AA\
\citep{aar-bommier77,aar-landi_d3_82}, and it should be also a
reasonable approximation for those of the He\,I 10830\,\AA\
multiplet which result from transitions between the terms 2s$^3$S and
2p$^3$P.

It is well-known that the He\,\uppercase{i} atom can be correctly
described by the L-S coupling scheme
\citep[e.g.,][]{aar-condon_shortley35}. The different $J$-levels are
grouped in terms with well defined values of the electronic angular
momentum $L$ and the spin $S$. The energy separation between the
$J$-levels inside each term is very small in comparison with the
energy difference between different terms.  Therefore, in addition to
the atomic polarization of each $J$-level (population imbalances and
coherences between its magnetic substates), it turns out to be
fundamental to allow for coherences between different $J$-levels
pertaining to the same term. However, coherences between different
$J$-levels pertaining to different terms can be safely neglected. In
conclusion, we can describe the atom via the formalism of the
multi-term atom (see Sections 7.5 and 7.6 of
\cite{aar-landi_landolfi04}).

The atomic level polarization is quantified by the multipole
components, $\rho^K_Q(J,J^{'})$, of the atomic density matrix
(\cite{aar-landi_landolfi04}). The number of real quantities required
to describe the excitation state of our multi-term model atom is
405. The helium atoms are illuminated by the (given) anisotropic
radiation field coming from the underlying solar photosphere
\citep{aar-pierce00}, whose properties are quantified by two spherical
tensors: $J^0_0$ and $J^2_0$ (see Eq. 5.164 in
\cite{aar-landi_landolfi04}).  They describe the mean intensity and
the ``degree of anisotropy'' of the radiation field, respectively. We
also consider that a magnetic field $\mathbf{B}$ is present, with
strength $B$, inclination $\theta_B$ with respect to the local
vertical and azimuth $\chi_B$.  The density matrix elements are
obtained by solving the statistical equilibrium equations \citep[see
Section 7.6a of][]{aar-landi_landolfi04}, which take into account the
effect of the radiative transitions in the presence of the Zeeman
splittings produced by the assumed magnetic field vector.

Once the elements of the atomic density matrix are known, the
coefficients of the emission vector and of the propagation matrix of
the Stokes-vector transfer equation can be directly calculated
\citep[see Section 7.6b of][]{aar-landi_landolfi04}. The next step is
to compute the emergent Stokes profiles by solving the Stokes-vector
transfer equation. To this aim, our code includes several options. The
simplest one assumes that the emergent Stokes profiles are simply
proportional to the corresponding emissivity.  This optically-thin
case is representative of the conditions found in some
prominences. The second option assumes that the He\,\uppercase{i} atoms
are located in a slab of constant physical properties characterized by
a non-negligible optical depth $\Delta \tau$ and that magneto-optical
effects are negligible (see Trujillo Bueno et al.\ 2005 for the
analytical expressions of the emergent Stokes profiles).  Finally, the
most general radiative transfer option is accomplished by using the
DELOPAR method \citep{aar-trujillo03}.

Following the previously described approach, we have developed an
efficient computer program that we have combined with an easy-to-use
interactive front-end.  It allows to interactively select the physical
effects to be taken into account, thus helping to achieve a fast and
reliable investigation of the influence of various physical parameters
on the emergent Stokes parameters \citep[for a first application,
see][]{aar-trujillo_helium07}. For instance, our code allows to
include or discard the effect of the magneto-optical terms and/or
stimulated emission processes. Furthermore, it is possible to carry
out the forward modeling including or neglecting the presence of
atomic polarization.  When atomic polarization is not taken into
account, the problem reduces to the simplified case of the Zeeman
effect, but with the positions and strengths of the $\sigma$ and $\pi$
components calculated within the framework of the incomplete
Paschen-Back effect theory. This is important for the He\,I
10830\,\AA\ multiplet because the linear Zeeman effect theory
underestimates the inferred magnetic field strength
\citep{aar-socas_navarro04}.  In the absence of atomic level
polarization, the emergent spectral line radiation is polarized due to
the non-zero magnetic splitting produced by the presence of the
magnetic field. For completeness, the Milne-Eddington solution of the
radiative transfer equation can also be used in this case.  By
default, the total Hamiltonian (including the fine structure and
magnetic contributions) is diagonalized numerically. However, our code
also permits to calculate the emergent Stokes profiles assuming the
linear Zeeman regime, which is useful in order to compare with those
calculated within the framework of incomplete Paschen-Back theory.

\section{The Stokes Inversion Option}
The previously described forward modeling code has been also combined
with a robust inversion scheme.  The inversion algorithm is based on
the minimization of a merit function that describes how well our model
reproduces the observed Stokes profiles.  This merit function is
chosen to be the standard $\chi^2$--function (least square
method). The minimization algorithm tries to find the value of the
model's parameters that lead to synthetic Stokes profiles in closest
agreement with the observed ones (i.e., to those that imply the
smallest value of the merit function). The model's parameters are the
magnetic field vector $\mathbf{B}$, the thermal velocity
$v_\mathrm{th}$, the macroscopic velocity $v_\mathrm{mac}$ and the
damping of the line $a$. If radiative transfer effects are taken into
account, this set is augmented with the optical depth of the slab
($\Delta \tau$).

The standard Levenberg-Marquardt (LM) procedure for the minimization
of the $\chi^2$ function has been used. The LM method is one of the
fastest and simplest available methods when the initial estimation of
the parameters is close to the correct minimum. On the contrary, like
with the majority of the standard numerical methods for function
minimization, its main drawback is that it can easily get trapped in a
local minimum of the $\chi^2$ function. A straightforward but time
consuming strategy to overcome this difficulty is to restart the
minimization process with different values of the initial parameters.

\begin{figure}
  \centering
  \includegraphics[width=\textwidth]{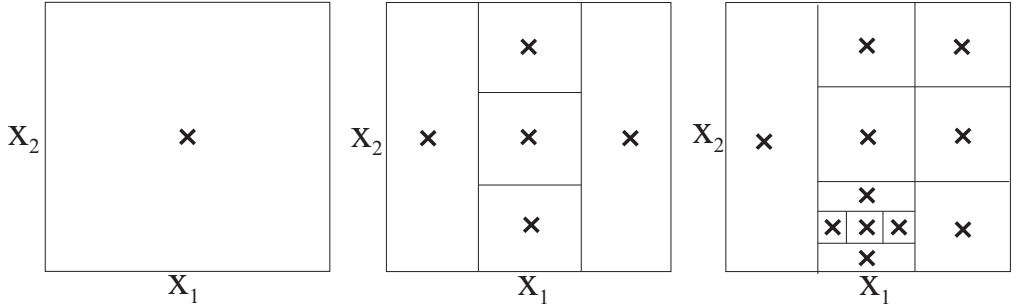}
  \caption[]{This figure illustrates the philosophy of the DIRECT method for searching 
for the region where the global minimum is 
located. In this case, we illustrate a case in two dimensions. After the evaluation of 
the function at some selected points inside each region, 
the DIRECT algorithm decides, using the Lipschitz condition, which rectangles can 
be further subdivided because they are potentially interesting. The method rapidly 
locates the region where the minimum is located.\label{asensio-fig:direct}
}
\end{figure}

In our case, this difficulty has been overcome by using global
optimization methods (GOM) that provide a good starting point. The
majority of this type of methods are based on stochastic optimization
techniques. The key idea is to efficiently sample the whole space of
parameters to find the global minimum of a given function. One of the
most promising methods is genetic optimization. In spite of the lack
of a convergence theorem, they perform quite well in practice
\citep[e.g.,][]{aar-Lagg04}.  In our case, we propose to apply instead
the DIRECT deterministic method that relies on a strong mathematical
basis. The name stands for ``DIviding RECTangles''
\citep{aar-Jones_DIRECT93} and the idea is to recursively sample parts
of the space of parameters, locating in each iteration the part of the
space where the global minimum is potentially located. The decision
algorithm is based on the assumption that the function is Lipschitz
continuous. For clarifying purposes, consider the case of a
one-dimensional function $f: M \to R$. It is said to be Lipschitz
continuous with constant $\alpha$ in an interval $M=[a,b]$ if:
\begin{equation}
|f(x)-f(x')| \leq \alpha |x-x'|, \qquad \forall x, x'\in M.
\end{equation}
In particular, a consequence of the previous equation is that the two
following equations have to be fulfilled $\forall x \in M$:
\begin{eqnarray}
f(x) &\geq& f(a) - \alpha (x-a) \\
f(x) &\geq& f(b) + \alpha (x-b) .
\end{eqnarray}
These two straight lines form a V-shape below the function $f(x)$
whose intersection provides a first estimation $x_1$ of the minimum of
the function. Repetition of this procedure in the two new subintervals
$[a,x_1]$ and $[x_1,b]$ leads to new estimations $x_i$ that converge
to the global minimum. Severe problems arise when generalizing this
method to higher dimensions because of the difficulty in estimating
the Lipschitz constant. Actually, the function is often not Lipschitz
continuous. The DIRECT algorithm overcomes these problems because it
does not require knowledge of the Lipschitz constant. It uses all
possible values of such a constant to determine if a region of the
space of parameters should be broken into subregions because of its
potential interest. Fig.~\ref{asensio-fig:direct} shows a schematic
illustration of the subdivision process for a function of two
parameters.

\begin{table}[t]
\begin{center}
\begin{tabular}{cccc}
\tableline\tableline
Step & Method & Free parameters & Stokes profiles \\
\tableline
1 & DIRECT & $v_\mathrm{th}$, $v_\mathrm{mac}$, $\Delta \tau$, $a$ & $I$ \\
2 & LM & $v_\mathrm{th}$, $v_\mathrm{mac}$, $\Delta \tau$, $a$ & $I$ \\
3 & DIRECT & $B$, $\theta_B$, $\chi_B$ & $I$, $Q$, $U$, $V$ \\
4 & LM & $B$, $\theta_B$, $\chi_B$ & $I$, $Q$, $U$, $V$ \\
\tableline
\end{tabular}
\end{center}
\caption[]{Scheme applied for the inversion of the Stokes profiles. The
two DIRECT cycles are used for obtaining an initial value of the
parameters as close to the global minimum as possible. This is
accomplished by the fast global convergence of DIRECT. Then, a few LM
iterations are sufficient to rapidly refine the value of the
minimum.\label{asensio-tab:inversion}}
\end{table}

One of the main drawbacks of deterministic global optimization methods
is that they present a very good global convergence but a poor local
convergence. Therefore, they quickly locate the region of the space of
parameters where the global minimum is located, but the refinement of
this solution takes a very long time. Consequently, the DIRECT method
is a perfect candidate for its application as an estimator of the
initial value of the parameters that are then rapidly refined by the
LM algorithm. Since the initial point is very close to the minimum,
the LM method, thanks to the quadratic convergence behavior, takes a
few iterations to rapidly converge towards the global minimum. We have
confirmed that this scheme reduces the number of evaluations of the
merit function. It is important to point out that, because the most
time-consuming part of the optimization procedure is the evaluation of
the complex forward problem (with the solution of the statistical
equilibrium equations and the radiative transfer), it is fundamental
to minimize the number of merit function evaluations.

A critical and fundamental point in the optimization of functions is
the stopping criterion.  In the few (unrealistic) cases where the
value of the function at the global minimum is known (noise-free
cases), it is possible to stop the convergence process when the
relative error is smaller than a fixed value. In the general case with
DIRECT, very good results are obtained when the hypervolume (in the
parameter space) where the algorithm is looking for the global minimum
has been reduced by a given factor, $f \approx 0.001$, with respect to
the original hypervolume. Another possible option is to stop after a
fixed number of evaluations of the merit function.

The full inversion scheme, shown schematically in Table
\ref{asensio-tab:inversion}, is started with the DIRECT method to
obtain a first estimation of the thermodynamical parameters by using
only Stokes $I$. After this initialization, some iterations of the LM
method are carried out to refine the initial values of the
thermodynamical parameters obtained in the previous step. Once the LM
method has converged, the inferred values of $v_\mathrm{th}$,
$v_\mathrm{mac}$, and $a$ (and $\Delta \tau$ when it applies) are kept
fixed. In a next step, the DIRECT method is used again for the
initialization of the magnetic field vector ($B$, $\theta_B$ and
$\chi_B$). According to our experience, the first estimation of the
magnetic field vector given by the DIRECT algorithm is typically very
close to the final solution. As a final step, some iterations of the
LM method are performed to refine the value of the magnetic field
strength, inclination and azimuth until reaching the global
optimum. Obviously, if any parameter is known in advance, we can keep
it fixed during the whole inversion process.

\begin{figure}
  \centering
  \includegraphics[width=\textwidth]{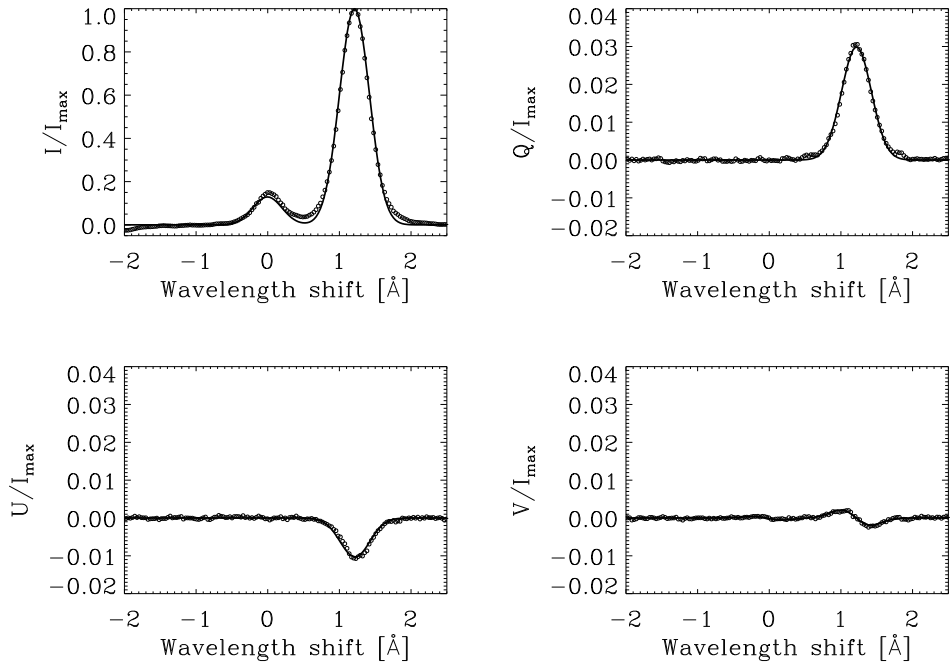}
  \caption[]{Example of the best theoretical fit obtained with the inversion code to the 
observed profiles presented in \citet{aar-merenda06}.
The emergent Stokes profiles have been obtained by assuming an optically thin 
plasma. The inferred magnetic field vector is given by 
$B=26.8$\,G, $\theta_B=25.5^\circ$ and $\chi_B=161^\circ$. Furthermore, the 
inferred thermal velocity is 
$v_\mathrm{th}=7.97$~km s$^{-1}$. All these values are in close agreement with 
the results obtained by \citet{aar-merenda06}.
The positive direction of Stokes $Q$ is the one parallel to the solar limb.\label{asensio-fig:prominence}
}
\end{figure}

\begin{figure}
  \centering
  \includegraphics[width=\textwidth]{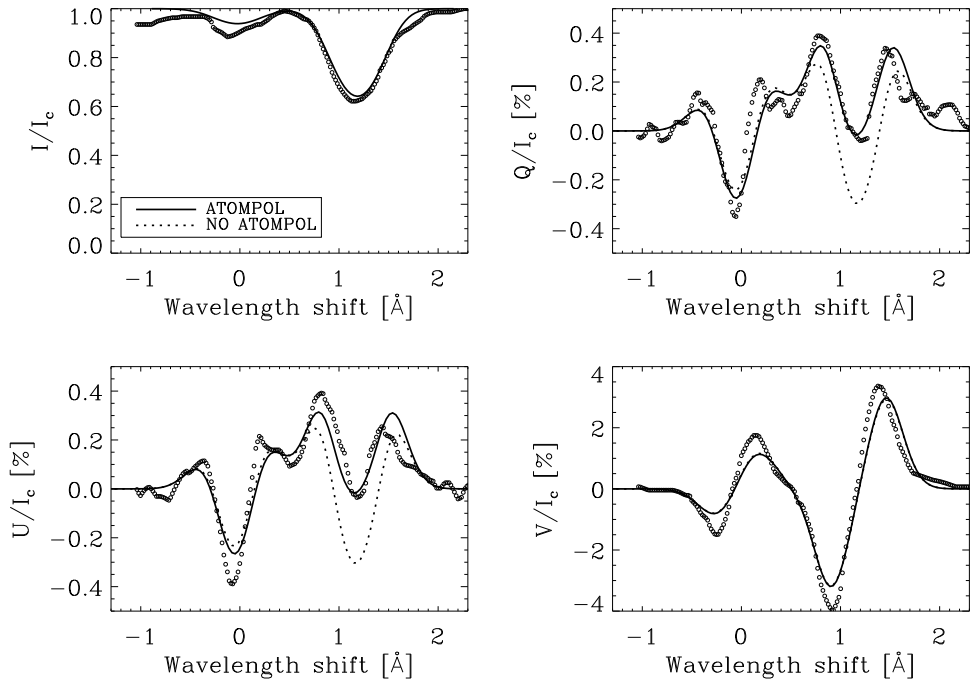}
  \caption[]{Emergent Stokes profiles of the He\,I 10830\,\AA\ 
multiplet in an emerging flux region. 
The circles represent the observations shown by \citet{aar-Lagg04} in their Fig.~2. The dotted lines present the best 
theoretical fit obtained neglecting the influence of atomic level polarization, which corresponds to
$B=1070$\,G, $\theta_B=86^{\circ}$ and $\chi_B=-160^{\circ}$. On the contrary, the solid lines show the best
fit when the effect of atomic level polarization is included, giving
$B=1009$\,G, $\theta_B=91^{\circ}$ and $\chi_B=-161^{\circ}$. The best fit to the 
observations is provided by the solid lines, which indicates the presence of atomic level polarization
in a relatively strong field region.\label{asensio-fig:emerging_flux}
}
\end{figure}

\section{Illustrative Applications}
Here we have only space to demonstrate the application of this new
plasma diagnostic tool to the case of a prominence and to that of an
emerging magnetic flux region.  For more details and applications see
Asensio Ramos \& Trujillo Bueno (in preparation).

\subsection{A polar crown prominence} 
For illustrative purposes, we have applied our inversion code to infer
the magnetic field vector of the polar crown prominence that generated
the Stokes profiles presented in Fig.~9 of \citet{aar-merenda06}. We
have assumed that the prominence plasma was optically thin. Our
inversion code was used to infer the value of the thermal velocity
$v_\mathrm{th}$ and the magnetic field vector
$(B,\theta_B,\chi_B)$. The height of the prominence atoms was fixed at
$h=20\arcsec$, the same value used by \citet{aar-merenda06}. After the four
steps summarized in Table \ref{asensio-tab:inversion}, we ended up
with a thermal velocity of $v_\mathrm{th}=7.97$~km s$^{-1}$ and a
magnetic field vector characterized by $B=26.8$\,G,
$\theta_B=25.5^\circ$, and $\chi_B=161.0^\circ$.  The inferred magnetic
field vector is in very good agreement with that obtained by
\citet{aar-merenda06}, namely $B=26$\,G, $\theta_B=25^\circ$, and
$\chi_B=160.5^\circ$.  Therefore, we fully support their conclusion of
nearly-vertical fields in the observed polar-crown prominence.

\subsection{An emerging Magnetic flux region}
As pointed out by \citet{aar-trujillo_helium07}, the modeling of the
emergent Stokes $Q$ and $U$ profiles of the He\,I 10830\,\AA\
multiplet should be done by taking into account the possible presence
of atomic level polarization, even for magnetic field strengths as
large as 1000\,G.  An example of a spectropolarimetric observation of
an emerging magnetic flux region is shown by the circles of
Fig.~\ref{asensio-fig:emerging_flux}.  The solid lines show the best
theoretical fit to these observations of \citet{aar-Lagg04}. Here, in
addition to the Zeeman effect, we took into account the influence of
atomic level polarization. The dotted lines neglect the atomic level
polarization that is induced by anisotropic radiation pumping in the
solar atmosphere. Our results indicate the presence of atomic level
polarization in a relatively strong field region (${\sim}$1000
G). However, it may be tranquilizing to point out that both inversions
of the observed profiles yield a similar magnetic field vector, in
spite of the fact that the corresponding theoretical fit is much
better for the case that includes atomic level polarization.

\acknowledgements
We would like to thank Egidio Landi Degl'Innocenti for sharing with 
us his deep knowledge on the physics of spectral line polarization. 
This research has been funded by the
Spanish Ministerio de Educaci\'on y Ciencia
through project AYA2004-05792.



\end{document}